\documentclass[a4paper,twocolumn,english,aps, pra]{revtex4-1}
\usepackage[latin9]{inputenc}
\usepackage{units}
\usepackage{textcomp}
\usepackage{amssymb}
\usepackage{graphicx}

\makeatletter

\newcommand*\LyXThinSpace{\,\hspace{0pt}}

\@ifundefined{textcolor}{}
{%
 \definecolor{BLACK}{gray}{0}
 \definecolor{WHITE}{gray}{1}
 \definecolor{RED}{rgb}{1,0,0}
 \definecolor{GREEN}{rgb}{0,1,0}
 \definecolor{BLUE}{rgb}{0,0,1}
 \definecolor{CYAN}{cmyk}{1,0,0,0}
 \definecolor{MAGENTA}{cmyk}{0,1,0,0}
 \definecolor{YELLOW}{cmyk}{0,0,1,0}
}

\usepackage{currfile}

\usepackage{babel}

\makeatother

\usepackage{babel}
\begin{document}

\title{Heisenberg Uncertainty Relation for Three Canonical Observables}

\author{Spiros Kechrimparis}

\email{sk864@york.ac.uk}

\author{Stefan Weigert}

\email{stefan.weigert@york.ac.uk}

\affiliation{Department of Mathematics, University of York, York YO10 5DD, United
Kingdom}

\date{Received 18 July 2014; published 15 December 2014}

\pacs{03.65.Ta,03.67.\textminus a}
\begin{abstract}
Uncertainty relations provide fundamental limits on what can be said
about the properties of quantum systems. For a quantum particle, the
commutation relation of position and momentum observables entails
Heisenberg's uncertainty relation. A third observable is presented
which satisfies canonical commutation relations with both position
and momentum. The resulting triple of pairwise canonical observables
gives rise to a Heisenberg uncertainty relation for the product of
three standard deviations. We derive the smallest possible value of
this bound and determine the specific squeezed state which saturates
the triple uncertainty relation. Quantum optical experiments are proposed
to verify our findings.

\global\long\def\kb#1#2{|#1\rangle\langle#2|}

\end{abstract}
\maketitle
\global\long\def\bk#1#2{\langle#1|#2\rangle}

\global\long\def\braket#1#2{\langle#1|#2\rangle}

\global\long\def\ket#1{|#1\rangle}

\global\long\def\bra#1{\langle#1|}

\global\long\def\c#1{\mathbb{C}^{#1}}

\global\long\def\abs#1{\mid#1\mid}

\global\long\def\avg#1{\langle#1\rangle}

\section{Introduction}

In quantum theory, two observables $\hat{p}$ and $\hat{q}$ are canonical
if they satisfy the commutation relation 
\begin{equation}
\left[\hat{p},\hat{q}\right]=\frac{\hbar}{i}\,,\label{eq: PairCommutationRelations}
\end{equation}
with the momentum and position of a particle being a well-known and
important example. The non-vanishing commutator expresses the incompatibility
of the Schrödinger pair $(\hat{p},\hat{q})$ of observables since
it imposes a lower limit on the product of their standard deviations,
namely 
\begin{equation}
\Delta q\,\Delta p\geq\frac{\hbar}{2}\,.\label{eq: PairUncertaintyRelation}
\end{equation}

In 1927, Heisenberg \cite{heisenberg27} analysed the hypothetical
observation of an individual electron with photons and concluded that
the product of the measurement errors should be governed by a relation
of the form (\ref{eq: PairUncertaintyRelation}). His proposal inspired
Kennard \cite{kennard27} and Weyl \cite{weyl28} to mathematically
derive Heisenberg's uncertainty relation, thereby turning it into
a constraint on measurement outcomes for an ensemble of identically
prepared systems. Schrödinger's \cite{schroedinger30} generalization
of (\ref{eq: PairUncertaintyRelation}) included a correlation term,
and Robertson \cite{robertson29,robertson34} derived a similar relation
for any two non-commuting Hermitean operators. Recently claimed violations
of (\ref{eq: PairUncertaintyRelation}) do not refer to Kennard and
Weyl's \emph{preparation} uncertainty relation but to Heisenberg's
\emph{error-disturbance} relation (cf. \cite{ozawa03,erhart+12,rozema+12}).
However, these claims have been criticized strongly \cite{busch+13,dressel+14}.

Uncertainty relations are now understood to provide fundamental limits
on what can be said about the properties of quantum systems. Imagine
measuring the standard deviations $\Delta p$ and $\Delta q$ separately
on two ensembles prepared in the same quantum state. Then, the bound
(\ref{eq: PairUncertaintyRelation}) does not allow one to simultaneously
attribute definite values to the observables $\hat{p}$ and $\hat{q}$.

In this paper, we will consider a \emph{Schrödinger triple} $(\hat{p},\hat{q},\hat{r})$
consisting of \emph{three} pairwise canonical\emph{ }observables \cite{weigert+08},
i.e. 
\begin{equation}
\left[\hat{p},\hat{q}\right]=\left[\hat{q},\hat{r}\right]=\left[\hat{r},\hat{p}\right]=\frac{\hbar}{i}\,,\label{eq: TripleCommutationRelations}
\end{equation}
and derive a \emph{triple} uncertainty relation. In a system of units
where both $\hat{p}$ and $\hat{q}$ carry physical dimensions of
$\sqrt{\hbar}$, the observable $\hat{r}$ is given by 
\begin{equation}
\hat{r}=-\hat{q}-\hat{p}\label{eq: constraint on pqr}
\end{equation}
which corresponds to a suitably \emph{rotated }and \emph{rescaled
}position operator $\hat{q}$. It is important to point out that any
Schrödinger triple for a quantum system with one degree of freedom
is unitarily equivalent to $(\hat{p},\hat{q},\hat{r})$; furthermore,
any such triple is maximal in the sense that there are no four observables
that equi-commute to $\hbar/i$ \cite{weigert14}. Therefore, the
algebraic structure defined by a Schrödinger triple $(\hat{p},\hat{q},\hat{r})$
is unique up to unitary transformations.

Given that (\ref{eq: PairCommutationRelations}) implies Heisenberg's
uncertainty relation (\ref{eq: PairUncertaintyRelation}), we wish
to determine the consequences of the commutation relations (\ref{eq: TripleCommutationRelations})
on the product of the \emph{three} uncertainties associated with a
Schrödinger triple $(\hat{p},\hat{q},\hat{r})$.

\section{Results}

We will establish the \emph{triple uncertainty relation} 
\begin{equation}
\Delta p\,\Delta q\,\Delta r\;\geq\left(\tau\,\frac{\hbar}{2}\right)^{\nicefrac{3}{2}}\,,\label{eq: TripleUncertainty}
\end{equation}
where the number $\tau$ is the \emph{triple constant} with value
\begin{equation}
\tau=\csc\left(\frac{2\pi}{3}\right)\equiv\sqrt{\frac{4}{3}}\simeq1.16\,.\label{eq: tripleconstant}
\end{equation}
The bound (\ref{eq: TripleUncertainty}) is found to be \emph{tight};
the state of minimal triple uncertainty is found to be a generalized
squeezed state, 
\begin{equation}
\ket{\Xi_{0}}=\hat{S}_{\frac{i}{4}\ln3}\,\ket 0\,,\label{eq:XI as squeezed state-1}
\end{equation}
being \emph{unique except for rigid translations in phase space.}
The operator $\hat{S}_{\frac{i}{4}\ln3}$$ $, defined in Eq.~(\ref{eq: real squeeze operator})
is a generalized squeezing operator: it generates the state $\ket{\Xi_{0}}$
by \emph{contracting} the standard coherent state $\ket 0$ (i.e.,
the ground state of a harmonic oscillator with unit mass and unit
frequency) along the main diagonal in phase space by an amount characterized
by $\ln\sqrt[4]{3}<1$, at the expense of a \emph{dilation} along
the minor diagonal.

\begin{figure}
\includegraphics[scale=0.5]{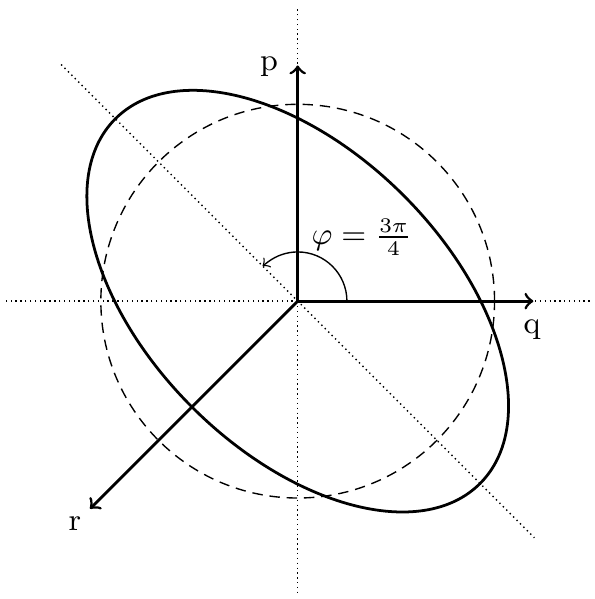}\protect\caption{\label{fig: contour plot}Phase-space contour lines of the Wigner
functions associated with the states $\protect\ket{\Xi_{0}}$ (full
line) and a standard coherent state $\protect\ket 0$ (dashed), respectively;
both lines enclose the same area. }
\end{figure}

To visualize this result, let us determine the Wigner function of
the state $\ket{\Xi_{0}}$ with position representation (cf. \cite{moller+96})
\begin{equation}
\bk q{\Xi_{0}}=\frac{1}{\sqrt[4]{\tau\pi}}\exp\left(-\frac{1}{2}e^{-i\frac{\pi}{6}}q^{2}\right)\,.\label{MinStR-1-1}
\end{equation}
Thus, its Wigner function associated with the state $\ket{\Xi_{0}}$
minimizing the triple uncertainty relation is found to be 
\begin{equation}
W_{\Xi_{0}}(q,p)=\frac{1}{\pi}\exp\left(-\frac{\tau}{\hbar}\left(q{}^{2}+p{}^{2}+qp\right)\right)\,,
\end{equation}
which is positive. Its phase-space contour line enclosing an area
of size $\hbar$, shown in Fig. \ref{fig: contour plot}, confirms
that we deal with a squeezed state aligned with the minor diagonal.

To appreciate the bound (\ref{eq: TripleUncertainty}), let us evaluate
the triple uncertainty $\Delta p\Delta q\Delta r$ in two instructive
cases. (i) Since the pairs $\left(\hat{p},\hat{q}\right)$, $\left(\hat{q},\hat{r}\right)$,
and $\left(\hat{r},\hat{p}\right)$ are canonical, the inequality
(\ref{eq: PairUncertaintyRelation})---as well as its generalization
due to Robertson and Schrödinger---applies to each of them implying
the lower bound 
\begin{equation}
\Delta p\,\Delta q\,\Delta r\geq\left(\frac{\hbar}{2}\right)^{\nicefrac{3}{2}}\,.\label{eq: TripleUncertaintyPairs}
\end{equation}
However, it remains open whether there is a state in which the triple
uncertainty saturates this bound. Our main result (\ref{eq: TripleUncertainty})
reveals that \emph{no} such state exists. (ii) In the vacuum $\ket 0$,
a coherent state with minimal pair uncertainty, the \emph{triple}
uncertainty takes the value 
\begin{equation}
\Delta p\,\Delta q\,\Delta r=\sqrt{2}\left(\frac{\hbar}{2}\right)^{\nicefrac{3}{2}}\,.\label{eq: TripleUncertaintyCoherentState}
\end{equation}
The factor of $\sqrt{2}$ in comparison with (\ref{eq: TripleUncertaintyPairs})
has an intuitive explanation: while the vacuum state $\ket 0$ successfully
minimizes the product $\Delta p\,\Delta q$, it does not simultaneously
minimize the uncertainty associated with the pairs $\left(\hat{q},\hat{r}\right)$
and $\left(\hat{r},\hat{p}\right)$. Thus, the minimum of the inequality
(\ref{eq: TripleUncertainty}) cannot be achieved by coherent states.

The observations (i) and (ii) suggest that the bound (\ref{eq: TripleUncertainty})
on the triple uncertainty is not an immediate consequence of Heisenberg's
inequality for canonical \emph{pairs}, Eq.~(\ref{eq: PairUncertaintyRelation}).
Furthermore, the invariance groups of the triple uncertainty relation,
of Heisenberg's uncertainty relation, and of the inequality by Schrödinger
and Robertson are different, because they depend on two, three and
four (cf. \cite{trifonov01}) continuous parameters, respectively.

\begin{figure}
\begin{centering}
\includegraphics[clip,scale=0.5]{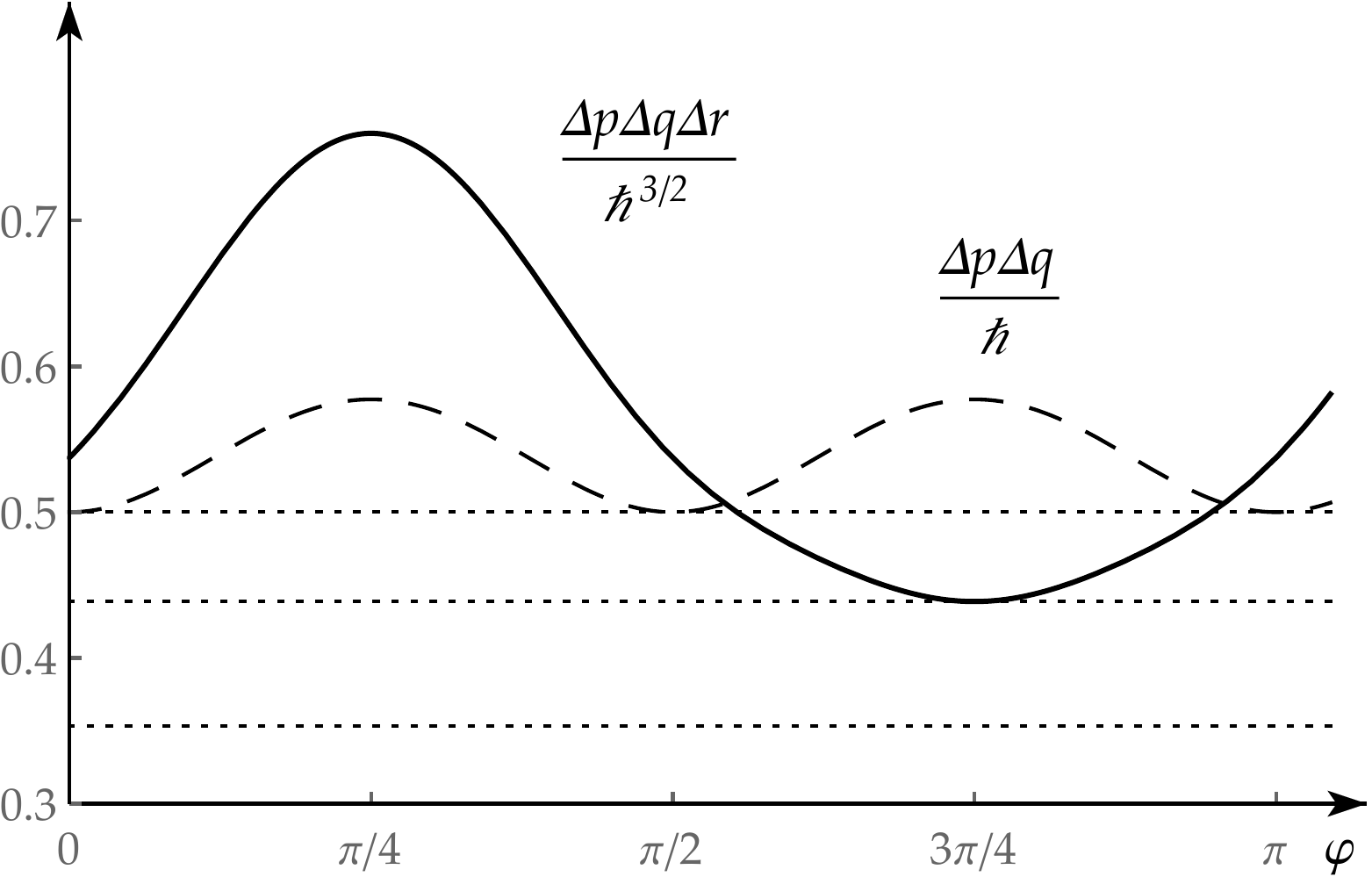} 
\par\end{centering}

\protect\caption{\label{fig: rotated triple variance-1} Dimensionless pair and triple
uncertainties for squeezed states with $\gamma=\ln\sqrt[4]{3}$, rotated
away from the position axis by an angle $\varphi\in[0,\pi]$ . The
pair uncertainty $\Delta p\Delta q$ starts out at its minimum value
of $1/2$ which is achieved again for $\varphi=\pi/2$ and $\varphi=\pi$
(dashed line). The triple uncertainty has period $\pi$, reaching
its minimum for $\varphi=3\pi/4$ for the state $\protect\ket{\Xi_{0}}$
(full line). The dotted lines (top to bottom) represent the bounds
(\ref{eq: PairUncertaintyRelation}), (\ref{eq: TripleUncertainty}),
and (\ref{eq: TripleUncertaintyPairs}), with values $\ensuremath{\nicefrac{1}{2}}$,
$\ensuremath{\left(\nicefrac{\tau}{2}\right)^{\nicefrac{3}{2}}}$,
and $\ensuremath{\left(\nicefrac{1}{2}\right)^{\nicefrac{3}{2}}}$.}
\end{figure}

\section{Threefold symmetry}

The commutation relations (\ref{eq: TripleCommutationRelations})
are invariant under the cyclic shift $\hat{p}\rightarrow\hat{q}\rightarrow\hat{r}\rightarrow\hat{p}$,
implemented by a unitary operator $\hat{Z}$, 
\begin{equation}
\hat{Z}\hat{p}\hat{Z}^{\dagger}=\hat{q}\,,\quad\hat{Z}\hat{q}\hat{Z}^{\dagger}=\hat{r}\,,\quad\hat{Z}\hat{r}\hat{Z}^{\dagger}=\hat{p}\,.\label{eq: CycleOperator-1}
\end{equation}
Note that the third equation follows from the other two equations.
The third power of $\hat{Z}$ obviously commutes with both $\hat{p}$
and $\hat{q}$ so it must be a scalar multiple of the identity, $\hat{Z}^{3}\propto\hat{\mathbb{I}}$.

To determine the operator $\hat{Z}$ we first note that its action
displayed in (\ref{eq: CycleOperator-1}) is achieved by a clockwise
rotation by $\pi/2$ in phase space followed by a gauge transformation
in the position basis: 
\begin{equation}
\hat{Z}=\exp\left(-\frac{i}{2\hbar}\hat{q}^{2}\right)\exp\left(-\frac{i\pi}{4\hbar}\left(\hat{p}^{2}+\hat{q}^{2}\right)\right)\,.\label{eq: C as a product-1}
\end{equation}
A Baker-Campbell-Hausdorff (BCH) calculation reexpresses this product
in terms of a single exponential: 
\begin{equation}
\hat{Z}=\exp\left(-i\frac{\pi}{3\hbar\sqrt{3}}\left(\hat{p}^{2}+\hat{q}^{2}+\hat{r}^{2}\right)\right)\,.\label{eq: combined C-1}
\end{equation}

The operator $\hat{Z}$ cycles the elements of the Schrödinger triple
$(\hat{p},\hat{q},\hat{r})$ just as a Fourier transform operator
swaps position and momentum of the Schrödinger pair $(\hat{p},\hat{q})$
(apart from a sign). If one introduces a unitarily equivalent symmetric
form of the Schrödinger triple with operators $(\hat{P},\hat{Q},\hat{R})$
associated with an equilateral triangle in phase space, the metaplectic
operator $\hat{Z}$ simply acts as a rotation by $2\pi/3$, i.e.,
as a fractional Fourier transform.

Furthermore, denoting the factors of $\hat{Z}$ in (\ref{eq: C as a product-1})
by $\hat{A}$ and $\hat{B}$ (with suitably chosen phase factors),
respectively, we find that $\hat{B}^{2}=\hat{\mathbb{I}}$ and $(\hat{A}\hat{B})^{3}\equiv\hat{Z}^{3}=\hat{\mathbb{I}}$.
These relations establish a direct link between the threefold symmetry
of the Schrödinger triple $(\hat{p},\hat{q},\hat{r})$ and the\emph{
modular group} $SL_{2}(\mathbb{Z})/\{\pm1\}$ which $\hat{A}$ and
$\hat{B}$ generate \cite{serre73}.

\section{Experiments\label{sec: ExperimentalVerification-1}}

To experimentally confirm the triple uncertainty relation (\ref{eq: TripleUncertainty}),
we propose an approach based on optical homodyne detection. We exploit
the fact that the state $\ket{\Xi_{0}}$ is a \emph{generalized coherent
state}, also known as a \emph{correlated coherent state} \cite{dodonov+80}:
such a state is obtained by squeezing the vacuum state $\ket 0$ along
the momentum axis followed by a suitable rotation in phase space.

The basic scheme for homodyne detection consists of a beam splitter,
photodetectors and a reference beam, called the \emph{local oscillator},
with which the signal is mixed; by adjusting the phase of the local
oscillator one can probe different directions in phase space. If $\theta$
is the phase of the local oscillator, a homodyne detector measures
the probablity distribution of the observable 
\begin{equation}
\hat{x}(\theta)=\frac{1}{\sqrt{2}}\left(a^{\dagger}e^{i\theta}+ae^{-i\theta}\right)=\hat{q}\cos\theta+\hat{p}\sin\theta\,
\end{equation}
along a line in phase space defined by the angle $\theta$; here $\hat{q}$
and $\hat{p}$ denote the quadratures of the photon field while the
operators $a^{\dagger}$ and $a$ create and annihilate single photons
\cite{welsch+09}; note that $\hat{r}\equiv\sqrt{2}\,\hat{x}(5\pi/4)$.

The probability distributions of the observables $\hat{q},\hat{p}$
and $\hat{r}$, corresponding to the angles $\theta=0$, $\pi/2$,
and $5\pi/4$, can be measured upon preparing a large ensemble of
the state $\ket{\Xi_{0}}$. The resulting product of their variances
may then be compared with the value of the tight bound given in Eq.~(\ref{eq: TripleUncertainty}).
Under rigid phase-space rotations of the triple $(\hat{q},\hat{p},\hat{r})$
by an angle $\varphi$ the triple uncertainty will vary as predicted
in Fig. \ref{fig: rotated triple variance-1} (full line). A related
experiment has been carried out successfully in order to directly
verify other Heisenberg- and Schrödinger-Robertson-type uncertainty
relations \cite{manko+09,bellini+2012}.

\section{Minimal triple uncertainty\label{sub: extremaltripleuncertainty} }

To determine the states which minimize the left-hand-side of Eq.~(\ref{eq: TripleUncertainty}),
need to evaluate it for all normalized states $\ket{\psi}\in{\cal H}$
of a quantum particle. To this end we introduce the \emph{uncertainty
functional} (cf. \cite{jackiw68}), 
\begin{equation}
J_{\lambda}[\psi]=\Delta_{p}[\psi]\,\Delta_{q}[\psi]\,\Delta_{r}[\psi]-\lambda(\braket{\psi}{\psi}-1)\,,\label{eq: TripleFunctional}
\end{equation}
using the standard deviations $\Delta_{x}[\psi]\equiv\Delta x$ $\equiv\left(\bra{\psi}\hat{x}^{2}\ket{\psi}\right.$$\left.-\bra{\psi}\hat{x}\ket{\psi}^{2}\right)^{\nicefrac{1}{2}}\,$,
$x=p,q,r$, while the term with Lagrange multiplier $\lambda$ takes
care of normalization. In a first step, we determine the extremals
of the functional $J_{\lambda}[\psi]$. Changing its argument from
$\ket{\psi}$ to the state $\ket{\psi}+\ket{\varepsilon}$, where
$\ket{\varepsilon}=\varepsilon\ket e$, with a normalized state $\ket e\in{\cal H}$
and a real parameter $\varepsilon\ll1$, leads to 
\begin{equation}
J_{\lambda}[\psi+\varepsilon]=J_{\lambda}[\psi]+\varepsilon J_{\lambda}^{(1)}[\psi]+\mathcal{O}(\varepsilon^{2})\,.\label{eq: ExpandJtoFirstOrder}
\end{equation}
The first-order variation $J_{\lambda}^{(1)}[\psi]$$ $ only vanishes
if $\ket{\psi}$ is an extremum of the functional $J_{\lambda}[\psi]$
or, equivalently, if 
\begin{equation}
\frac{1}{3}\left(\frac{\left(\hat{p}-\avg{\hat{p}}\right)^{2}}{\Delta_{p}^{2}}+\frac{\left(\hat{q}-\avg{\hat{q}}\right)^{2}}{\Delta_{q}^{2}}+\frac{\left(\hat{r}-\avg{\hat{r}}\right)^{2}}{\Delta_{r}^{2}}\right)\ket{\psi}=\ket{\psi}\label{eq: TripleStationaryCondition}
\end{equation}
holds, which follows from generalizing a direct computation which
had been carried out in\emph{ }\cite{weigert96} to determine the
extremals\emph{ }of the product $\Delta p\,\Delta q$.

Eq.~(\ref{eq: TripleStationaryCondition}) is non-linear in the unknown
state $\ket{\psi}$ due to the expectation values $\avg{\hat{p}},\Delta_{p}^{2}$,
etc. Its solutions can be found by initially treating these expectation
values as constants to be determined only later in a self-consistent
way. The unitary operator $\hat{U}_{\alpha,b,\gamma}$$=$$\hat{T}_{\alpha}\hat{G}_{b}\hat{S}_{\gamma}$
transforms the left-hand side of (\ref{eq: TripleStationaryCondition}),
which is quadratic in $\hat{p}$ and $\hat{q}$, into a standard harmonic-oscillator
Hamiltonian, 
\begin{equation}
\frac{1}{2}\left(\hat{p}^{2}+\hat{q}^{2}\right)\ket{\psi_{\alpha,b,\gamma}}=\frac{3}{2c}\ket{\psi_{\alpha,b,\gamma}}\,,\label{eq: gammaExtremals}
\end{equation}
where $\ket{\psi_{\alpha,b,\gamma}}\equiv\hat{U}_{\alpha,b,\gamma}^{\dagger}\ket{\psi}$,
and $c$ is a real constant. The unitary $\hat{U}_{\alpha,b,\gamma}$
consists of a \emph{rigid phase-space translation} by $\alpha\equiv(q_{0}+ip_{0})/\sqrt{2\hbar}\in\mathbb{C}$,
\begin{equation}
\hat{T}_{\alpha}=\exp\left[i\left(p_{0}\hat{q}-q_{0}\hat{p}\right)/\hbar\right]\,,\label{eq: Translation operator}
\end{equation}
followed by a \emph{gauge transformation} in the momentum basis 
\begin{equation}
\hat{G}_{b}=\exp\left(ib\hat{p}^{2}/2\hbar\right)\,,\quad b\in\mathbb{R}\,,\label{eq: GaugeOperator}
\end{equation}
and a \emph{squeezing transformation}, 
\begin{equation}
\hat{S}_{\gamma}\equiv\exp[i\gamma(\hat{q}\hat{p}+\hat{p}\hat{q})/2\hbar]\,,\quad\gamma\in\mathbb{R}\,.\label{eq: real squeeze operator}
\end{equation}

According to (\ref{eq: gammaExtremals}), the states $\ket{\psi_{\alpha,b,\gamma}}$
coincide with the eigenstates $\ket n,n\in\mathbb{N}_{0}$, of a harmonic
oscillator with unit mass and frequency, 
\begin{equation}
\ket{n;\alpha,b,\gamma}\equiv\hat{T}_{\alpha}\hat{G}_{b}\hat{S}_{\gamma}\ket n\,,\qquad n\in\mathbb{N}_{0}\,,\label{eq: AllStationaryStates(preliminary)}
\end{equation}
where we have suppressed irrelevant constant phase factors; for consistency,
the quantity $3/2c$ in (\ref{eq: gammaExtremals}) must only take
the values $\hbar(n+1/2)$ for $n\in\mathbb{N}_{0}$, as a direct
but lengthy calculation confirms. The parameters $b$ and $\gamma$
must take specific values for (\ref{eq: gammaExtremals}) to hold,
namely 
\begin{equation}
b=\frac{1}{2}\qquad\mbox{and}\qquad\gamma=\frac{1}{2}\ln\tau\,;\label{eq: b and gamma}
\end{equation}
we will denote the restricted set of states obtained from Eq.~(\ref{eq: AllStationaryStates(preliminary)})
by $\ket{n;\alpha}$. There are no constraints on the parameter $\alpha$,
which means that we are free to displace the states $\ket n$ in phase
space without affecting the values of the variances. The variances
of the observables $\hat{p},\hat{q},$ and $\hat{r}$ are found to
be equal, taking the value 
\begin{equation}
\Delta_{x}^{2}[n;\alpha]=\tau\hbar\left(n+\frac{1}{2}\right)\,,\quad x=p,q,r\,,\label{eq:squaredpqrVariancesFinal}
\end{equation}
with the triple constant $\tau$ introduced in (\ref{eq: tripleconstant}).
Inserting these results into Eq.~(\ref{eq: TripleStationaryCondition})
we find that 
\begin{equation}
\frac{1}{3}\left(\hat{p}^{2}+\hat{q}^{2}+\hat{r}^{2}\right)\ket{n;\alpha}=\tau\hbar\left(n+\frac{1}{2}\right)\,\ket{n;\alpha}\,,\label{eq: TripleStationaryConditionFinal}
\end{equation}
where 
\begin{equation}
\ket{n;\alpha}=\hat{T}_{\alpha}\hat{G}_{\frac{1}{2}}\hat{S}_{\frac{1}{2}\ln\tau}\ket n\,,\quad n\in\mathbb{N}_{0}\,,\alpha\in\mathbb{C}\,.\label{eq: AllStationaryStates(final)}
\end{equation}
For each value of $\alpha$, the \emph{extremals} of the uncertainty
functional (\ref{eq: TripleFunctional}) form a complete set of orthonormal
states, 
\begin{equation}
\sum_{n=0}^{\infty}\kb{n;\alpha}{n;\alpha}=\mathbb{I}\,,\label{eq: oscillator resolution}
\end{equation}
since the set of states $\left\{ \ket n\right\} $$ $ has this property.

At its extremals the uncertainty functional (\ref{eq: TripleFunctional})
takes the values 
\begin{equation}
J_{\lambda}[n;\alpha]=\left[\tau\,\hbar\left(n+\frac{1}{2}\right)\right]^{\nicefrac{3}{2}}\,,\quad n\in\mathbb{N}_{0}\,,\label{eq: extemal values of J-1}
\end{equation}
according to Eq.~(\ref{eq:squaredpqrVariancesFinal}), with the minimum
occuring for $n=0$. Thus, the two-parameter family of states $\ket{0;\alpha},\alpha\in\mathbb{C}$,
which we will denote by 
\begin{equation}
\ket{\Xi_{\alpha}}=\hat{T}_{\alpha}\left(\hat{G}_{\frac{1}{2}}\hat{S}_{\frac{1}{2}\ln\tau}\ket 0\right)\,,\label{eq: Xi_alpha-1}
\end{equation}
minimize the triple uncertainty relation (\ref{eq: TripleUncertainty}).

The states $\ket{\Xi_{\alpha}}$ are \emph{displaced generalized squeezed}
states, with a squeezing direction along a line different from the
position or momentum axes. To show this, it is sufficient to consider
the state $\ket{\Xi_{0}}$, which satisfies (\ref{eq: TripleStationaryConditionFinal})
with $n\equiv0$ and $\alpha\equiv0$. The product of unitaries in
(\ref{eq: Xi_alpha-1}) acting on the vacuum $\ket 0$ is easily understood
if one rewrites it using the identity 
\begin{equation}
\hat{G_{b}}\hat{S}_{\gamma}=\hat{S}_{\xi}\hat{R}_{\varphi}\,,\label{eq: GS=00003D00003DSR-1}
\end{equation}
where the unitary $\hat{R}_{\varphi}=\exp(i\varphi a^{\dagger}a)$
is a counterclockwise rotation by $\varphi$ in phase space while
the operator 
\begin{equation}
\hat{S}_{\xi}=\exp\left[\left(\overline{\xi}a^{2}-\xi a^{\dagger2}\right)/2\right]\,,\quad\xi=\gamma e^{i\theta}\,,\gamma>0\,,\label{eq: general squeeze operator-1}
\end{equation}
generalizes $\hat{S}_{\gamma}$ in (\ref{eq: real squeeze operator})
by allowing for squeezing along a line with inclination $\theta/2$;
the annihilation operator and its adjoint $a^{\dagger}$ are defined
by $a=\left(\hat{q}+i\hat{p}\right)/\sqrt{2\hbar}$. Another standard
BCH calculation (using result from Sec. 6 of \cite{wilcox67}) reveals
that the values $\xi=(i/4)\ln3$ and $\varphi=-\pi/12$ turn Eq.~(\ref{eq: GS=00003D00003DSR-1})
into an identity for the values of $b$ and $\gamma$ given in (\ref{eq: b and gamma}).
This confirms that the state of minimal triple uncertainty is the
generalized squeezed state given in (\ref{eq:XI as squeezed state-1}).

\section{Summary and Discussion}

We have established a tight inequality (\ref{eq: TripleUncertainty})
for the triple uncertainty associated with a Schrödinger triple $(\hat{p},\hat{q},\hat{r})$
of pairwise canonical observables. Ignoring rigid translations in
phase space, there is only one state $\ket{\Xi_{0}}$ which minimizes
the triple uncertainty, shown in Eq.~(\ref{eq: Xi_alpha-1}). The
state $\ket{\Xi_{0}}$ is an eigenstate of the operator $\hat{Z}$
in (\ref{eq: combined C-1}) which describes the fundamental threefold
cyclic symmetry of the Schrödinger triple $(\hat{p},\hat{q},\hat{r})$.
Conceptually, the triple uncertainty and the one derived by Schrödinger
and Robertson are linked because both incorporate the correlation
operator $(\hat{p}\hat{q}+\hat{q}\hat{p})/2$, be it explicitly or
indirectly via the expression $\hat{r}^{2}$.

The smallest possible value of the product $\Delta p\Delta q\Delta r$
is noticably \emph{larger} than the unachievable value $\left(\hbar/2\right)^{\nicefrac{3}{2}}$,
which follows from inequality (\ref{eq: PairUncertaintyRelation})
applied to each of the Schrödinger pairs $(\hat{p},\hat{q})$, $(\hat{q},\hat{r})$,
and $(\hat{r},\hat{p})$. At the same time, the true minimum \emph{undercuts}
the value of the triple uncertainty in the vacuum state $\ket 0$
by more than $10\%$ {[}cf. Eq. (\ref{eq: TripleUncertaintyCoherentState}){]}.
The experimental verification of these results is within reach of
current quantum optical technology.

The results obtained in this paper add another dimension to the problem
of earlier attempts to obtain uncertainty relations for more than
two observables. In 1934, Robertson studied constraints which follow
from the positive semi-definiteness of the covariance matrix for $N$
observables \cite{robertson34} but the resulting inequality trivializes
for an odd number of observables. Shirokov obtained another inequality
\cite{shirokov04} which contains little information about the canonical
triple considered here.

The result for a Schrödinger triple obtained here suggests conceptually
important generalizations. A tight bound for an \emph{additive} uncertainty
relation associated with the operators $(\hat{p},\hat{q},\hat{r})$
is easily established by a similar approach: the inequality

\begin{equation}
(\Delta p)^{2}+(\Delta q)^{2}+(\Delta r)^{2}\geq\tau\,\frac{3\hbar}{2}\label{eq: triple sum inequality}
\end{equation}
is saturated only by the state $\ket{\Xi_{0}}$ in (\ref{eq: Xi_alpha-1}),
ignoring irrelevant rigid phase-space translations. This observation
clashes with the relation between the \emph{additive }and the\emph{
multiplicative }uncertainty relations for Schrödinger \emph{pairs}
$(\hat{p},\hat{q})$. According to \cite{trifonov01} the states saturating
the inequality $(\Delta p)^{2}+(\Delta q)^{2}\geq\hbar$ are a proper
subset of those minimizing Heisenberg's \emph{product} inequality
(\ref{eq: PairUncertaintyRelation}).

Finally, an uncertainty relation for pairs of canonical observables
also exists for the Shannon entropies $S_{p}$ and $S_{q}$ of their
probability distributions \cite{hirschman57,bialnicki+75}. We conjecture
that the relation $S_{p}+S_{q}+S_{r}\geq(3/2)\ln(\tau e\pi)$ holds
for the Schrödinger triple $(\hat{p},\hat{q},\hat{r})$, the minimum
being achieved by the state $\ket{\Xi_{0}}$. This bound is tighter
than $(3/2)\ln(e\pi)$, the value which follows from applying the
bound $\ln(e\pi)$ for pairwise entropies to the triple.
\begin{acknowledgments}
S.K. is supported via the act ``Scholarship Programme of SSF by the
procedure of individual assessment, of 2011--12'' by resource of
the Operational Programme for Education and Lifelong Learning of the
ESF and of the NSF, 2007--2013. S.W. would like to thank V.I. Man'ko
for an instructive discussion on aspects of the experimental realization.
The authors thank J. Biniok, T. Bullock, P. Busch and R. Colbeck for
comments on drafts of this manuscript. \end{acknowledgments}

\end{document}